\documentclass[aps,prd,reprint,superscriptaddress,nofootinbib, longbibliography,preprintnumbers]{revtex4-1}

\usepackage{amsmath,amssymb, amsthm,amstext,comment,lipsum,mathtools,placeins}
\usepackage{natbib}
\usepackage{graphicx}
\usepackage{color}
\usepackage{array, enumerate}
\usepackage{bm}
\usepackage{multirow}
\usepackage[breaklinks,colorlinks,citecolor=blue]{hyperref}
\usepackage{braket}
\usepackage{txfonts}
\usepackage[normalem]{ulem}

\def\be{\begin{equation}}
\def\ee{\end{equation}}

\def\apjl{Astroph.J.Lett.}

\def\araa{ARA\&A}

\newcommand{\dd}{{\rm d}}
\newcommand{\bS}{ \boldsymbol{\mathcal{S}} }
\newcommand{\bx}{\boldsymbol{x}}
\newcommand{\bTh}{\boldsymbol{\Theta}}
\newcommand{\bt}{\boldsymbol{t}}
\newcommand{\bn}{\boldsymbol{n}}

\newcommand{\Mc}{\mathcal{M}}

\begin{document}

\preprint{DESY-24-137}
\title{Non-Gaussian Statistics of Nanohertz Stochastic Gravitational Waves}

\author{Xiao Xue}
\email{xxue@ifae.es}
\affiliation{Institut de Física d’Altes Energies (IFAE), The Barcelona Institute of Science and Technology, Campus UAB, 08193 Bellaterra (Barcelona), Spain}
\affiliation{II. Institute of Theoretical Physics, Universit\"{a}t Hamburg, 22761 Hamburg, Germany}
\affiliation{Deutsches Elektronen-Synchrotron DESY, Notkestr. 85, 22607 Hamburg, Germany}
\author{Zhen Pan}
\email{zhpan@sjtu.edu.cn}
\affiliation{Tsung-Dao Lee Institute, Shanghai Jiao-Tong University, Shanghai, 520 Shengrong Road, 201210, People’s Republic of China}
\affiliation{School of Physics \& Astronomy, Shanghai Jiao-Tong University, Shanghai, 800 Dongchuan Road, 200240, People’s Republic of China}
\author{Liang Dai}
\email{liangdai@berkeley.edu}
\affiliation{Department of Physics, University of California, 366 Physics North MC 7300, Berkeley, CA. 94720, USA}

\begin{abstract}

Multiple pulsar timing arrays (PTAs) have recently reported evidence for nHz stochastic gravitational wave background (SGWB), stimulating intensive discussions about its physical origin.
In principle, the sources may be either supermassive black hole binaries (SMBHBs) or processes in the early Universe. One key difference between the two lies in the statistics of the SGWB frequency power spectrum.
In particular, the often assumed Gaussian random SGWB does not accurately describe the distribution of the collective SMBHB emission.
This work presents a semianalytical framework for calculating the non-Gaussian statistics of SGWB power expected from SMBHBs.
We find that (a) wave interference between individual SMBHBs with indistinguishable observed frequencies and (b) the Poisson fluctuation of the source numbers, together shape the non-Gaussian statistics. 
Implementing the non-Gaussian statistics developed in this work, we investigate the sensitivity of current and future PTA datasets in distinguishing the origin of the SGWB through non-Gaussian information. Additionally, we find an interesting approximation of the non-Gaussian statistics, which has implications for accurately and practically treating non-Gaussianity in PTA Bayesian analyses.

\end{abstract}
\date{\today}

\maketitle
\section{Introduction}
The recent detection of nanohertz (nHz) stochastic gravitational wave background (SGWB) by pulsar timing arrays (PTAs)  \cite{NANOGrav:2023gor,NANOGrav:2023hde,InternationalPulsarTimingArray:2023mzf, Reardon:2023gzh,Zic:2023gta, Antoniadis:2023rey,Antoniadis:2023utw, Xu:2023wog} has inspired intensive discussions of its astrophysical implications. Supermassive black hole binaries (SMBHBs) are a promising source of SGWB, from which we may infer the abundance and evolution history of the cosmic SMBHB population \cite{NANOGrav:2023hfp,Bi:2023tib,sato2023nanograv,Ellis:2023dgf,Ellis:2024qpv,Ellis:2024wdh,Padmanabhan:2024nvv,Liepold:2024woa}.
On the other hand, various early-Universe processes have been speculated to be alternative nHz SGWB sources (see \cite{NANOGrav:2023hvm} for a summary). 

In principle, SMBHBs and early Universe sources can be distinguished by the different statistical properties of the SGWB they produce. SMBHBs at redshifts $z < 1$ contribute significantly to the SGWB at nanohertz frequencies, characterized by strong signal strength due to their proximity, and they emit mostly monochromatic GWs over decades of observation. In contrast, early Universe sources, produced at high redshift, are heavily redshifted and contribute minimally due to their larger distance. These sources usually emit over a broad frequency range. A much larger number of sources would be required for early Universe sources to contribute to the same signal power as SMBHBs at the same observed frequency. Thus, it is usually assumed that the latter produces a Gaussian and isotropic SGWB, while 
 the former is certainly non-Gaussian due to Poissonian fluctuations of a finite number of SMBHBs and may show random power anisotropy if there are nearby loud SMBHBs.
 Along this line, there have been some efforts towards
 measuring the spectral variance beyond Gaussian fluctuations \cite{Roebber:2015iva,Ellis:2023oxs,Lamb:2024gbh,Bernardo:2024uiq,Agazie:2024jbf,Sardesai:2024rdd,Raidal:2024odr},
 or anisotropies of the nHz SGWB \cite{Hotinli:2019tpc,Taylor:2020zpk,Pol:2022sjn,NANOGrav:2023tcn,Lemke:2024cdu,Depta:2024ykq,Gardiner:2023zzr,Yang:2024mqz}.
 In these analyses, spectral variance is usually quantified or parametrized based on SMBHB population synthesis. 
Specially, refs.~\cite{Ellis:2023owy,Sato-Polito:2024lew} derived the probability distribution of the characteristic strain $P(h_c^2)$ given a SMBHB population model, which enabled a quantitative constraint on SMBHB population models using the PTA data \cite{Sato-Polito:2024lew}. For the study of anisotropy, forecasts are made based on SMBHB population simulations,
 and measurements from the PTA data are performed under the simplifying assumption of a Gaussian random SGWB.
 Spectral variance and anisotropy are two different manifestations of Poisson fluctuations of a finite number of SMBHBs,
  in the intrinsic (e.g., GW amplitude and frequency) and  extrinsic (e.g., sky localization) parameter spaces, respectively. 
 Focusing on a different aspect, 
 Allen {\it et al}. \cite{Allen:2022dzg,Allen:2024rqk} calculated the 
 variance of the Hellings-Downs correlation \cite{Hellings-Downs} for a simple population of SMBHBs, 
 taking into account interference between GW sources of overlapping frequencies.
 They demonstrated two sources of variance: one arising from the finite number of pulsar pairs being used, and another cosmic variance due to interference effects.

As summarized above, different aspects of nHz SGWB sourced by a finite number of SMBHBs
are related. Both spectral variance and anisotropy originate from Poissonian fluctuation in the number of SMBHBs
and are affected by interference, and
deviation from the Hellings-Downs correlation can be result from interference 
\cite{Wu:2024xkp,Allen:2022dzg,Allen:2024rqk,Romano:2023zhb} or spatial anisotropy \cite{Cornish:2013aba, Taylor:2013esa,Konstandin:2024fyo}.
In previous works, each aspect has only been investigated separately, assuming other aspects are known or independent.
Here, we consider a unified framework for dealing with Poissonian fluctuations
in the number of SMBHBs and GW interference. 
 The major challenge is efficiently computing the non-Gaussian probability distribution $P(\delta z)$ for general SMBHB population models, where $\delta z$ is observed photon redshift from a given single pulsar in the network.

This work demonstrates an efficient method for computing $P(\delta z)$ that accounts for the Poisson statistics and interference.
The key is to use the cumulant  generating function for $\delta z$ (see Sec~\ref{sec:compound_poisson} for details).
As a result, we find that interference is the major source of variance in $|\delta z|^2$. At the same time, 
the non-Gaussianity originates from
the Poisson fluctuation in the source number  (see Fig.~\ref{fig:Violin}). With the non-Gaussian statistics $P(\delta z)$ developed in this work, we
estimate the likelihood ratio against the conventionally used Gaussian statistics, and find that the current PTA dataset is capable of finding substantial evidence of non-Gaussian statistics (Fig.~\ref{fig:lambdaLR}).
However, we find narrowing down the population model parameters challenging due to the significant statistical variance caused by 
interference (Fig.~\ref{fig:parameter_estimation}) given the current PTA sensitivity. As we will see, a more extended observation period and a lower timing noise level expected from future PTA data will enable substantially improved constraints on population model parameters.
Under current PTA sensitivity, we find the Gaussian statistics is still a good approximation in 
inferring the strain power spectrum $h_c^2(f)$ of the SGWB from SMBHBs (Fig.~\ref{fig:PLfit}).
As PTA data of lower timing noise accumulates in the foreseeable future, we find the  Gaussian statistics 
will eventually bias the inference of $h_c^2(f)$ (Fig.~\ref{fig:PLfit}).

 This paper is organized as follows. In Sec~\ref{sec:compound_poisson}, we briefly  introduce compound Poisson 
 statistics and the corresponding cumulant generating function (CGF) are key to computing the exact PTA signal distribution $P(\delta z)$ for SMBHB population models. 
 In Sec~\ref{sec:pop}, we introduce a parametrized form of the SMBHB population model to be considered in this work, 
 and provide details about calculating the redshift distribution $P(\delta z)$. Building on these,
 we perform a likelihood ratio test comparing non-Gaussian and Gaussian statistics.
 Concluding remarks will be given in Sec~\ref{sec:conclusions}. Throughout this work, we adopt a flat $\Lambda$ cold dark matter cosmology with $\Omega_{\rm m}=0.3$, $\Omega_{\Lambda}=0.7$, $H_0 = 70\ {\rm km}\, {\rm s}^{-1}\,{\rm Mpc}^{-1}$.

\section{Compound Poisson Statistics}
\label{sec:compound_poisson}

In this section, we develop the mathematical formalism for computing the probability density function (PDF) of the PTA observable, $P(\delta z)$. The basic idea was also explored in Ref.~\cite{Sato-Polito:2024lew}.
Assuming that individual gravitational wave sources are monochromatic over the observational period of PTAs, each source can be characterized by its current observed frequency $f_{\rm GW}$ and another set of $M$ parameters collectively referred to as $\bTh$. The following differential distribution can describe a general source population
\begin{equation}
     \frac{\dd^{1+M} \overline{N}}{\dd^M\bTh\, \dd \ln f}\ .
     \label{eq:gen_population}
\end{equation}
The average source number in a logarithmic frequency interval $\Delta \ln f$ and in a multidimensional parameter-space volume element $\Delta^M \bTh$ is thus given by $\Delta \overline{N}$.

The actual source number $\Delta N$ is expected to be a random number that fluctuates around its expectation value $\langle\Delta N\rangle = \Delta\overline{N}$. It follows Poisson statistics, i.e., 
\begin{equation}
    \Delta N\sim {\rm Pois}(\Delta \overline{N})\ .
\end{equation}
Source numbers in nonoverlapping volumes of the parameter space are independent random numbers.

For any signal $s(\bTh,f)$ that obeys the superposition principle, the population-summed signal $\mathcal{S}$ is a weighted sum over the entire source parameter space: 
\begin{equation}
    \mathcal{S} = \sum_n s_n\,\Delta N_n\ ,\label{eq:weighted_sum}
\end{equation}
where we divide the source parameter space into individual blocks indexed by $n$, and $s_n = s(\bTh_n,\,f^n_{\rm GW})$ is the corresponding weight. Linearity of Eq.~(\ref{eq:weighted_sum}) suggests that the total signal $\mathcal{S}$ is drawn from a compound Poisson distribution. In the special case where the weight $s(\bTh,f)$ is a constant, the total signal also follows a Poisson distribution. Unfortunately, we are not aware of an analytical expression for the PDF of $\mathcal{S}$, $P(\mathcal{S})$, for the case of general weights $s(\bTh,\,f)$.

Instead, we pursue a numerical method to evaluate the PDF $P(\mathcal{S})$ that makes use of  the CGF for $\mathcal{S}$, defined as
\begin{equation}
    K_{\mathcal S}(t)=\ln\left\langle \exp(i\, \mathcal{S}\,t) \right\rangle\ , \label{eq:K_sum}
\end{equation}
where $t$ is the variable conjugate to $\mathcal{S}$ and is defined on the real axis from $-\infty$ to $\infty$, and the notation $\langle \cdots \rangle$ stands for the statistical average. Since $\Delta N_n$'s are statistically independent of each other, Eq.~(\ref{eq:K_sum}) can be evaluated as follows
\begin{equation}
\begin{aligned}
    K_{\mathcal S}(t) 
    & = \sum_n\,\ln \left\langle \exp\left(i\,s_n\, \Delta N_n\,t\right) \right\rangle= \sum_n\, \Delta \overline N_n\,\left( e^{i\,s_n\,t}-1 \right)\ ,\label{eq:K_discrete}
    \end{aligned}
\end{equation}
where the last step follows from the CGF of Poisson distribution. Eq.~(\ref{eq:K_discrete}) then allows us to numerically evaluate the PDF of $\mathcal S$ through an inverse Fourier transformation 

\begin{equation}
    P(\mathcal{S}|\bTh,\,f)  = \frac{1}{2\pi}\int_{-\infty}^{+\infty} \dd t \, \exp\left(
    i\,\mathcal{S}\,t + [K_{\mathcal{S}}(t)]^*
    \right)\ ,\label{eq:fourier}
\end{equation}
where $[\cdots]^*$ stands for complex conjugation. The PDF evaluated in this way is correctly normalized.

If the source parameter space is continuously parametrized by $(\bTh,\,f)$, then Eq.~(\ref{eq:K_discrete}) can be revised to the appropriate continuous limit, where summation is replaced by integration
\begin{equation}
    \sum_n\,\Delta N_n \longrightarrow{} \int_V \dd\ln f\,\dd^M\bTh\, \frac{\dd^{1+M}\overline{N}}{\dd^M\bTh\,\dd\ln f}\ .
\end{equation}
The CGF converges in this limit, giving a unique answer for the PDF of the signal $\mathcal{S}$. 

It is straightforward to generalize the above framework to the case where a set of $\mathcal{N}$ different signals ($\mathcal{N}>1$), collectively denoted as $\bS$, are measured. The corresponding CGF depends on a set of $\mathcal{N}$ conjugate variables, which we collectively refer to as $\bt$. The CGF is given by
\begin{equation}
\begin{aligned}
    K_{\bS}(\bt) &= \sum_n\,\ln\left\langle \exp\left(i\, \boldsymbol{s}_n \cdot \bt\,\Delta N_n \right)\right\rangle = \sum_n\,\Delta \overline{N}_n\,\left(e^{i\,\boldsymbol{s}_n\cdot \bt}-1\right)\ , \label{eq:K_multi}
\end{aligned}
\end{equation}
and the multivariate PDF, in principle, can be found from the multidimensional Fourier transformation
\begin{equation}
     P(\bS|\,\bTh , \, f) = \frac{1}{(2\pi)^{\mathcal{N}}}\int \dd^\mathcal{N} \bt\, \exp\left(
     i\,\bS\cdot\bt+[K_{\bS}(\bt)]^*
     \right)\ .
\end{equation}

If an observable $\bS$ is also contaminated by noise $\bn$, and if the noise is statistically uncorrelated with the signal, then the total measurement $\bS +\bn$ has a CGF
\begin{align}
    K_{\bS+\bn}(\bt) = K_{\bS}(\bt) + K_{\bn}(\bt)\ ,\label{eq:combined_K}
\end{align}
where $K_{\bn}(\bt)$ is the CGF for the noise $\bn$. The corresponding PDF is essentially the likelihood function for obtaining data $\bS+\bn$ given a model with model parameters $\{\bTh , \varsigma\}$:
\begin{align}
    P(\bS+\bn|\,\bTh , \varsigma )\ ,
\end{align}
where $\varsigma$ is a set of parameters that characterize the noise. Based on the above general results following the probability theory, we will develop a framework to perform Bayesian inference of GW source population parameters.

\section{Non-Gaussian statistics}
\label{sec:pop}

In Sec~\ref{subsec:pop}, we will first introduce a convenient parametrized form for the SMBHB population following \cite{sato2023nanograv},
\be 
\frac{\dd^3\overline{N}}{\dd \ln f\,\dd\log_{10}\Mc\,\dd z}(\bTh )\ ,
\ee 
This gives the SMBHB number density in the phase space defined jointly by the GW frequency $f$, the binary chirp mass $\Mc$, and the cosmological redshift $z$.
We then introduce some basics of the PTA observable, redshift of photons $\delta z$ propagating from a pulsar to the earth
under the influence of GWs (Sec~\ref{subsec:redshift}), and the details for computing the non-Gaussian redshift PDF, $P(\delta z)$, given a SMBHB population model (Sec~\ref{subsec:pdf}).
With the redshift PDF, we compare the non-Gaussian statistics with the conventionally used Gaussian statistics in Sec~\ref{subsec:test}.

\subsection{SMBHB population}
\label{subsec:pop}

Supermassive black holes (SMBHs) are commonly found at the centers of their host galaxies. Galaxy mergers inevitably 
bring multiple SMBHs into the same postmerger galaxy. 
SMBHBs are the most promising sources for the nHz SGWB observed at PTAs,
though the detailed processes for the formation and mergers of SMBHBs still need to be fully understood.
The mass of the central SMBH is empirically known to correlate strongly with the velocity dispersion in the host galaxy bulge. This is the famous $M_{\rm BH}-\sigma$ relation \cite{Ferrarese2000,Kormendy2013,mcconnell2013revisiting}. 
We model the mean $M_{\rm BH}-\sigma$ relation, together with the expected scatter around it, with the following log-normal probability distribution
\begin{widetext}
\begin{align}
    P(\log_{10}M_{\rm BH}|
    \sigma) =\frac{1}{\sqrt{2\pi}\,\epsilon_0}\exp\left\{-\frac{1}{2\,\epsilon_0^2}
    \left[\log_{10}\frac{M_{\rm BH}}{M_{\odot}}-\log_{10}10^{a_{\bullet}}\left(\frac{\sigma}{200{\rm km}\,{\rm s}^{-1}}\right)^{b_\bullet}\right]^2\right\}\ ,
\end{align}
\end{widetext}
where $a_{\bullet}$ and $b_{\bullet}$ are constants and $\epsilon_0$ quantifies the intrinsic scatter. On the other hand, the velocity dispersion function (VDF) of the galaxy bulge is
parametrized as 
\begin{equation}
    \begin{aligned}
        \phi(\sigma) = \frac{\phi_*\,\beta}{\sigma_*} \left[ \frac{\sigma}{\sigma_*}\right]^{\alpha-1}\frac{e^{-(\sigma/\sigma_*)^{\beta}}}{\Gamma(\alpha/\beta)}\ .
    \end{aligned}
\end{equation}
Here $\alpha$, $\beta$ are dimensionless constants, $\sigma_*$ characterizes the turnover of the velocity dispersion, and $\phi_*$ is a normalization constant which also sets the total number of the galaxies per ${\rm Mpc}^{3}$. Taking these into account, the SMBH mass function is
\begin{equation}
\begin{aligned}
\frac{\dd n}{\dd \log_{10}M_{\rm BH}} = \int \dd\sigma\, P\left(\log_{10}\,M_{\rm BH}|\sigma\right)\phi(\sigma)\ ,
\end{aligned}
\end{equation}
where $n$ is the total number of black holes per ${\rm Mpc}^3$.

We assume that all SMBHBs in the PTA band are circular inspirals. This assumption is expected to be true when GW radiation dominates the hardening of the binary orbit, as GW emission tends to circularize the orbit. For circular inspirals, the chirp mass $\Mc$ and source redshift $z$ determine the observed GW amplitude. Chirp mass $\Mc$ is defined as
\begin{equation}
    \begin{aligned}
        \Mc = \eta^{3/5}M_{\rm BH},\qquad \eta \equiv \frac{q}{(1+q)^2}\ ,
    \end{aligned}
\end{equation}
where $q<1$ is the binary mass ratio. We express the population model as follows
\begin{equation}
    \frac{\dd n}{\dd \log_{10}\Mc} = \int_{q_{\rm min}}^1 \dd q \,p_q(q)\,\frac{\dd n }{\dd \log_{10}M_{\rm BH}}\ ,
\end{equation}
$p_q(q) = \mathcal{N}_q\,q^{\delta} $ and $\mathcal{N}_q$ are the normalization constants. The above expression assumes one merger per galaxy.

Only SMBHBs located on our past light cone are observed through the SGWB. The relevant comoving volume is given by $\dd V_c/\dd t$
\begin{equation}
    \frac{\dd V_c}{\dd t_r} = \frac{\dd V_c/\dd z}{\dd t_r /\dd z}=4\pi c\,d_L(z)^2\,(1+z)\ ,
\end{equation}
where $t_r$ is the proper time of a comoving observer in Friedmann-Robertson-Walker cosmology, and $d_L(z) = (c(1+z)/H_0)\,\int_0^z\,\dd z'/E(z')$ is the luminosity distance to redshift $z$, where $E(z) = \sqrt{(1+z)^3\Omega_{\rm m} + \Omega_{\Lambda}}$. Assuming the redshift dependence is separable, the SMBHB population can be written as \cite{sesana2008stochastic}
\begin{equation}
    \frac{\dd^3\overline{N}}{\dd \log_{10}\Mc\,\dd z\,\dd\ln f_r} = p_z(z)\,\frac{\dd n}{\dd \log_{10}\Mc}\,\frac{\dd t_r}{\dd \ln f_r}\,\frac{\dd V_c}{\dd t_r}\ .
    \end{equation}
where $f_r=(1+z)f$ is GW frequency in the rest frame of the SMBHB barycenter, $p_z(z) = \mathcal{N}_z\,z^{\gamma}\,e^{-z/z_*}$, 
$\dd t_r/\dd \ln f_r$ is determined by the binary orbital hardening rate,
$z_*$ is the turnover redshift, and $\mathcal{N}_z$ is a normalization constant. 
We adopt the simplifying assumption that the spectral shape of the mass function is independent of redshift $z$. 
We introduce $\xi$ to the frequency dependence
\begin{equation}
\begin{aligned}
    &\frac{\dd t_r}{ \dd \ln f_r} = \frac{5}{96\, (G\Mc/c^3)^{5/3}\,(\pi f_{\rm yr})^{8/3}(f_r/f_{\rm yr})^{8/3-\xi}}\ ,\label{eq:MBHB_dtdlnf}
    \end{aligned}
\end{equation}
especially $\xi=0$ if the hardening is driven by GW emission only. To summarize, in our parametrized SMBHB population model, $\{\epsilon_0,\,a_{\bullet},\,b_{\bullet}\}$ characterize the $M_{\rm BH}-\sigma$ relation,
$\{\phi_*,\,\sigma_*,\,\alpha,\,\beta\}$ characterize the VDF, and $\{\delta,\,\gamma,\,z_*,\,\xi\}$ characterize the SMBHB distribution with respect to binary mass ratio, redshift, and frequency.

\begin{figure}[t!]
    \centering
    \includegraphics[width=0.95\linewidth]{ 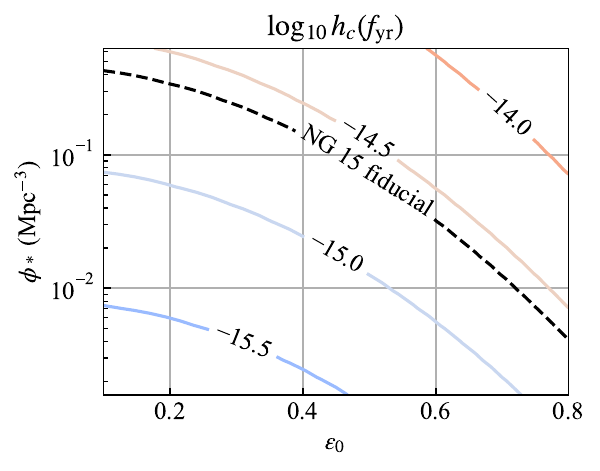}
    \caption{The contours of $\log_{10}(h_c)(f_{\rm yr})$ on $\phi_*-\epsilon_0$ plane. The parameters consistent with the 15-year NANOGrav data are marked as the black dashed line. We fix all the other parameters at their fiducial values (given in the text).}
    \label{fig:phi0_epsilon0}
\end{figure}

Following Ref.~\cite{sato2023nanograv}, we set fiducial values for the population parameters: $\sigma_*=159.6\ {\rm km }\,{\rm s}^{-1}$, $\{\alpha,\,\beta,;\,a_{\bullet},\,b_{\bullet};\,\gamma,\,z_*;\,\delta;\,\xi\}=\{0.41,\,2.59;\,8.32,\,5.64;\,0.3,\,0.5;\,-1;\,0 \} $. We leave $\phi_*$ and $\epsilon_0$ as free parameters. These choices are justified by studies on the velocity distribution function \cite{bernardi2010galaxy}, on the $M-\sigma$ relation \cite{mcconnell2013revisiting}, and SMBHB population simulations \cite{kelley2017massive}. Furthermore, in this work, we consider an expanded population model that accounts for source dependence on extra binary parameters in addition to $\Mc$ and $z$, 
\begin{equation}
    {\bf\Lambda} = \{\log_{10}\Mc,z\}\cup\{\cos\theta,\phi;\,\cos\iota,\psi;\,\varphi\}\ ,
\end{equation}
where $\iota$ is the binary orbit inclination, $\psi$ is the polarization angle, $\theta$ and $\phi$ parameterize the source's sky location, and $\varphi$ is a binary orbital phase constant. It is reasonable to assume that binaries have no preference regarding spatial location and orbital orientation. Therefore, the extended population model is
\begin{equation}
    \begin{aligned}
\frac{\dd^8\overline{N}}{\dd \ln f_{r}\,\dd^7 {\bf\Lambda}}
&= \frac{1}{32\pi^3}\frac{\dd^3\overline{N}}{\dd \ln f_{ r}\,\dd\log_{10}\Mc\,\dd z}\ .
\end{aligned}
\end{equation}

Next, we will look into the intensity of GWs from the circular SMBHBs. 
For a slowly-evolving binary with chirp mass $\Mc$ and redshift $z$, the GW tensor at the observer $\boldsymbol{x}=0$ can be written as 
\be 
h_{ab}(t,0) = h_+(t,0)\,\epsilon^+_{ab}+
h_\times(t,0)\,\epsilon^\times_{ab}\ ,
\ee 
where the polarization tensor $\epsilon^{+,\times}_{ab}(\theta,\phi,\psi)$ depends on the source sky location $(\theta,\phi)$ and the source polarization angle $\psi$ \cite{arzoumanian2023nanograv}.
The two polarization states have amplitudes
\be
\begin{aligned}
&h_+(t,0) = h_0\,\frac{1+\cos^2\iota}{2}\,\cos(2\pi f t+\varphi)\ , \\
&h_\times(t,0) = h_0\,\cos\iota\,\sin(2\pi f t+\varphi)\ ,
\end{aligned}
\ee 
where the dimensionless amplitude is
\be 
h_0(f, \Mc, z) = \frac{4c\,(\pi f_{\rm yr})^{2/3}\,(f_r/f_{\rm yr})^{2/3}\,(G\Mc/c^3)^{5/3}}{d_L(z)}.
\ee 
A commonly used quantity is the characteristic strain of the GWs, the expected value of which from a SMBHB population can be written as \cite{Phinney:2001di,sesana2008stochastic}
\begin{multline}\label{eq:hc2}
    h_c^2(f) = \int_0^{+\infty} \dd z \int_{-\infty}^{+\infty} \dd\log_{\rm 10} \Mc \,\langle  h_+^2+h_\times^2\rangle_{\cos\iota,\varphi}\\ \times\,\frac{\dd^3\overline{N}}{\dd \log_{10}\Mc\,\dd z\,\dd\ln f_r}\ ,
\end{multline}
where
\be \label{eq:hp22}
\langle h_+^2+h_\times^2\rangle_{\cos\iota,\varphi} =\frac{2}{5}\,h_0^2,
\ee
is the GW amplitude squared averaged over random binary orbital inclination and phase constant.
In Fig.~\ref{fig:phi0_epsilon0}, we show contours of $h_c(f_{\rm yr})$ on the $\phi_*$--$\epsilon_0$ plane, as well as the specific contour that corresponds to the best-fit value from the 15-year NANOGrav data \cite{NANOGrav:2023gor}, $h_c(f_{\rm yr}) = 2.4\times 10^{-15}$.

\subsection{Pulsar redshift}
\label{subsec:redshift}

As radio photons travel from a pulsar to the Earth, the arrival time is perturbed by the SGWB.
For a planar GW propagating in the direction $\hat{\boldsymbol{\Omega}}$ (with $-\hat{\boldsymbol{\Omega}}$ being the direction pointing from the Earth to the GW source),
 the GW tensor is generally written as
 \be 
h_{ab}(t, \boldsymbol{x}) = h_+(t-\boldsymbol{x}\cdot\hat{\boldsymbol{\Omega}})\epsilon^+_{ab}+
h_\times(t-\boldsymbol{x}\cdot\hat{\boldsymbol{\Omega}})\epsilon^\times_{ab}\ .
\ee 
Considering a pulsar with a sky location $\hat{\boldsymbol{p}}$ and at a distance $L$ from the earth, 
the observed photon redshift (or the fractional change in apparent pulsar spin period) for the observer at the coordinate origin $\boldsymbol{x}=0$ is \cite{Hellings-Downs}
\be\label{eq:delta_zt}
\begin{aligned}
    \delta z(t) &= \delta z_{\rm E} - \delta z_{\rm P} \\ 
    &= \frac{1}{2} \frac{\hat{p}^a\hat{p}^b}{1+\hat{\boldsymbol{\Omega}}\cdot \hat{\boldsymbol{p}}}
    \,\left[h_{ab}(t,\,0) - h_{ab}(t-L,\,L\,\hat{\boldsymbol{p} })\right]\ .
\end{aligned} 
\ee
Here, the subscripts E and P stand for the Earth and the pulsar terms.
For a slowly evolving SMBHB where the GW frequency evolution during the PTA observational period $T$ is negligible (with $2\pi\,\dot f\, T < T^{-1}$),
the two polarization components in the Earth term are in the form
\be
\begin{aligned}
&h_+(t, 0) = h_0\,\frac{1+\cos^2\iota}{2}\,\cos(2\pi\,f_{\rm E}\,t+\varphi)\ , \\
&h_\times(t, 0) = h_0\,\cos\iota\,\sin(2\pi\,f_{\rm E}\,t+\varphi)\ ,
\end{aligned}
\ee 
where the GW amplitude and the frequency $f_{\rm E}$ are taken as a constant during the observational period $(0,T)$.
Similarly, we have for the pulsar term 
\be
\begin{aligned}
&h_+(t-L,\,L\,\hat{\boldsymbol{p}}) = h_0\,\frac{1+\cos^2\iota}{2}\,\cos(2\pi\,f_{\rm P}\,t+\varphi-\Delta)\ , \\
&h_\times(t-L,\,L\,\hat{\boldsymbol{p}}) = h_0\,\cos\iota\,\sin(2\pi\,f_{\rm P}\,t+\varphi-\Delta)\ ,
\end{aligned}
\ee
where the phase difference $\Delta =c^{-1}\int_0^L 2\pi f(t)  {\rm d}t \ (1+\hat{\boldsymbol{\Omega}}\cdot \hat{\boldsymbol{p}})\approx \pi\,(f_{\rm E}+f_{\rm P})\,L\,(1+\hat{\boldsymbol{\Omega}}\cdot \hat{\boldsymbol{p}})/c$,
and $f_{\rm P}$ is also taken as a constant during the observational period $(0,T)$. 
We note that for a given pulsar, the frequency difference between the Earth term and the pulsar term, $f_{\rm E}-f_{\rm P}$, can be significant and measurable if $2\pi\,\dot f\,L /c> T^{-1}$. For SMBHB at nHz frequency, $f_{\rm E}$ and $f_{\rm P}$ are indistinguishable.

In practice, the PTA data analysis is more straightforward in the frequency domain, where the redshifts $\delta z(f)$ in different frequency bins are statistically uncorrelated.
We first divide the relevant frequency range into bins, with central frequencies $f_n$  ($n=1,...,n_{\rm max}$) and equal bin sizes $T^{-1}$.
The GW amplitude from a SMBHB in a frequency bin of central frequency $f_n$ is 
\begin{equation}
\begin{aligned}
        h_{ab}(f_n;f,\bTh)=&T\,W(f-f_n)\,h_0(f,\Mc,z)\,e^{i\varphi}\\&\,\,\times\left[
        i\,\frac{1+\cos^2\iota}{2}\,\epsilon^{+}_{ab} + \cos\iota\,\epsilon^{\times}_{ab}
        \right]\ ,
\end{aligned}  
\end{equation}
where $f$ is the observed GW frequency from the binary, 
and $W$ is the top-hat window function
 \begin{align}
     W(f-f_n) = 
     \begin{cases}
         1,& f_n-1/2T< f < f_n+1/2T\\
         0,& \text{otherwise} \ .
     \end{cases}
 \end{align}

As Eq.~(\ref{eq:delta_zt}) shows, two terms contribute to the redshift of a photon emitted from a pulsar and received on the Earth. However, it is widely accepted that only the Earth term exhibits interpulsar correlation, namely the Hellings-Downs curve \cite{hellings1983upper}. Here, we assume that we successfully subtracted the Earth term.
Taking the Earth term
\begin{align}
    \delta z_{\rm E}(f) = -\frac{1}{2\,T}\frac{p_a\,h_{ab}(\bx_{\rm E},f) \,p_b}{1+\hat{\boldsymbol{\Omega}}\cdot\hat{\boldsymbol{p}}}\ .
\end{align}
The expression above is dimensionless by introducing the prefactor $T^{-1}$. 
 Without loss of generality, we take $\hat{\boldsymbol{p}}=(0,0,1)$ for simplicity, then the real and imaginary components of $\delta z_{\rm E}$ are
\begin{equation}\label{eq:delta_z_reim}
    \begin{aligned}
        &\delta z_{\rm E,Re}^{\rm circ} = h_0\,\lambda_z(\theta,\iota,\psi,\varphi
        ),\\
        &\delta z_{\rm E,Im}^{\rm circ} = h_0\,\lambda_z(\theta,\iota,\psi,\varphi
       +\pi/2 ),
    \end{aligned}
\end{equation}
where
\begin{equation}
\begin{multlined}
        \lambda_{z}(\theta,\iota,\psi,\varphi) = \sin^2\frac{\theta}{2}\,\Bigg[\left(\frac{1+\cos\iota^2}{2}\right) \, \cos (2 \psi )\,\sin \varphi \\- \cos\iota \, \sin (2 \psi )\,\cos \varphi \Bigg]\ .
\end{multlined}
\end{equation}
For a population of SMBHBs, the total redshift is therefore 
\be 
\delta z_{\rm E} = \sum_n \Delta N_n\  (\delta z_{{\rm E}}^{\rm circ})_n\ ,\label{eq:interference}
\ee 
where we divide the source parameter space into individual
blocks indexed by $n$.
The distribution  $P(\lambda_z)$ is evaluated using the Monte Carlo method  by uniformly sampling $\cos\theta$, $\cos\iota$, $\psi$ and $\varphi$,
and we store numerical values for further use. The mean value of any arbitrary function $F(\lambda_z)$ is equivalent to the following integration 
\begin{widetext}
    \begin{equation}\label{eq:P_lambda}
    \begin{aligned}
        \langle F(\lambda_z)\rangle :=\int \dd\lambda_z\, P(\lambda_z)\,F(\lambda_z) = \frac{1}{32\,\pi^3}\int_{-1}^{1}\dd\cos\theta \int_{-1}^{1}\dd\cos\iota \int_{0}^{2\pi}\dd\phi \int_{0}^{2\pi}\dd\psi\int_{0}^{2\pi}\dd\varphi \,F(\lambda_z(\theta,\iota,\psi,\varphi))\ .
    \end{aligned}
\end{equation}
\end{widetext}
Then it is straightforward to find $ \langle \lambda_z\rangle=0$ and $ \langle \lambda_z^2\rangle =1/15$.

\begin{figure}[t!]
    \centering    
    \includegraphics[width=0.98\linewidth]{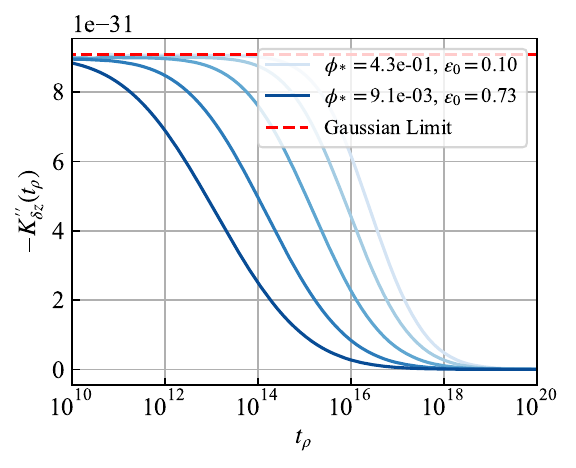}
    \includegraphics[width=0.98\linewidth]{ 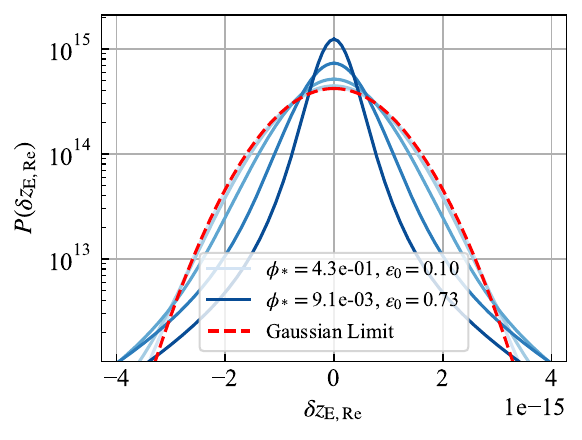}
    \caption{(Top) the second derivative of the CGF. (Bottom) the PDF of the observed redshift $\delta z_{\rm E,Re}$. Note that all PDFs in this plot share the same variance, $\langle \delta z_{\rm E,Re}^2\rangle$. We consider $f=10$ nHz with a bandwidth $\Delta f=2$ nHz.  Curves of lighter colors correspond to SMBHB populations with more SMBHBs but lower black hole masses. As the number of individual SMBHB increases, the CGF and the PDF approach the Gaussian limit. } 
    \label{fig:CGF_Krho}
\end{figure}

If a SMBHB is slowly evolving, with the Earth term and the pulsar term contributing to the same frequency bin, then we find 
\begin{equation}
\begin{aligned}
        \delta z^{\rm circ}&=\delta z_{\rm E}^{\rm circ}-\delta z_{\rm P}^{\rm circ} \\
        &= h_0 \left[ 
        \tilde\lambda_z(\theta,\iota,\psi,\varphi,\Delta) + i \tilde\lambda_z(\theta,\iota,\psi,\varphi+\pi/2,\Delta)\right]\ ,
\end{aligned}
\end{equation}
where
\be 
    \tilde\lambda_z(\theta, \iota, \psi, \varphi, \Delta) 
= 2\sin\left(\Delta/2\right) \lambda_z(\theta, \iota, \psi, \varphi -\Delta/2)\ .
\ee 
It is clear that the pulsar term further increases the variance of the redshift. The calculation of the redshift distribution, in this case, is completely parallel to the previous case with the replacement $\lambda_z(\theta, \iota, \psi, \varphi) \rightarrow \tilde \lambda_z(\theta, \iota, \psi, \varphi, \Delta)$. 
In the next subsection, we will focus on the PDF of the redshift $P(\delta z_{\rm E}(f))$. 

\subsection{Distribution of pulsar redshift}
\label{subsec:pdf}

Using the method developed in Sec~\ref{sec:compound_poisson}, 
we can now calculate the PDF of the redshift $P(\delta z_{\rm E}(f))$ given a SMBHB population model. 
We first assemble the real and imaginary parts into a vector $\boldsymbol{s}=(\delta z_{\rm E,Re}^{\rm circ}, \delta z_{\rm E,Im}^{\rm circ})$. 
The total signal is a superposition of many individual sources
and the conjugate variables also into a vector $\bt=(t_{\rm Re}, t_{\rm Im})$,
then  
\begin{equation}
\begin{aligned}
    \boldsymbol{s} \cdot \bt = \delta z_{\rm E,Re}^{\rm circ}\,t_{\rm Re} + \delta z_{\rm E,Im}^{\rm circ}\,t_{\rm Im} = t_{\rho}\,h_0\, \lambda_z(\theta,\iota,\psi,\varphi-t_{\varphi})\ ,\label{eq:trho}
\end{aligned}
\end{equation}
where $t_{\rm Re}$ and $t_{\rm Im}$ are real variables conjugate to $\delta z_{\rm E,Re}^{\rm circ}$ and $\delta z_{\rm E,Im}^{\rm circ}$, $ t_{\varphi}=\arctan(t_{\rm Im}/t_{\rm Re})$, $t_{\rho}=\sqrt{t_{\rm Re}^2 + t_{\rm Im}^2}$. 
Because of the uniform distribution of $\varphi$, the CGF does not depend on the variable $t_{\varphi}$. The two-dimensional CGF turns out to be 
\begin{equation}
\begin{aligned}
    K_{\delta z}(t_{\rho}) = \Delta\ln f\,\iiint \dd \log_{10}\Mc\,\dd z \,\dd \lambda_z \, P(\lambda_z) \\
    \times\left[\cos\left( t_{\rho}h_0\lambda_z \right) -1 \right]\frac{\dd^3 \overline{N}}{\dd \log_{10}\Mc\,\dd z\,\dd \ln f_r}\ .\label{eq:K_trtI}
\end{aligned}
\end{equation}
We find that $K_{\delta z}(t_{\rm Re},t_{\rm Im})=K_{\delta z}(t_\rho)$ is a real function because $P(\lambda_z)$ is even,
and $K_{\delta z}(t_{\rm Re},t_{\rm Im})$ only depends on the magnitude of the conjugate vector $t_\rho$.

The cumulants of the redshift can be inferred from the CGF $K_{\delta z}$.
First, it is easy to verify that all odd order of cumulants  vanish because $P(\lambda_z)$ is even,
and the higher and even order of cumulants, namely $\langle\delta z_{\rm E,Re}^{2k}\rangle$ with $k > 1$ do not exist because of the divergence at the nearby end (cosmology redshift $z\rightarrow 0$) of the SMBHB population, which implies that the corresponding PDF of $\delta z$ is a heavy-tailed distribution. The only nonzero and finite cumulant equals the variance of the redshift  $\delta z_{\rm E,Re}$ and $\delta z_{\rm E,Im}$, with
\begin{equation}\label{eq:variance}
    \begin{aligned}
    \langle \delta z_{\rm E,Re}^2 \rangle = \langle \delta z_{\rm E,Im}^2 \rangle  =  -K^{\prime\prime}_{\delta z}(0)\ .
    \end{aligned}
\end{equation}

In particular, $K^{\prime\prime}_{\delta z}(t_{\rho})$ is free of divergence at the nearby end of the SMBHB population:
\begin{equation}
\begin{aligned}
    K^{\prime\prime}_{\delta z}(t_{\rho}) = -\Delta\ln f\iiint \dd \log_{10}\Mc\dd z\, \dd \lambda_z \, P(\lambda_z) \\
    \times h_0^2\,\lambda_z^2\,\cos\left( t_{\rho}h_0\lambda_z \right) \,\frac{\dd^3 \overline{N}}{\dd \log_{10}\Mc \,\dd z\, \dd \ln f_r}\ .\label{eq:K_trtI}
\end{aligned}
\end{equation}
where it is straightforward to find $0\leq |K^{\prime\prime}_{\delta z}(t_{\rho}) / K^{\prime\prime}_{\delta z}(0)| \leq 1$ using Cauchy-Schwarz inequality,
and $ K^{\prime\prime}(t_\rho\to +\infty )=0$ due to the fast oscillatory integrand. 

\begin{figure*}[t]
    \centering
    \includegraphics[width=0.48\linewidth]{ 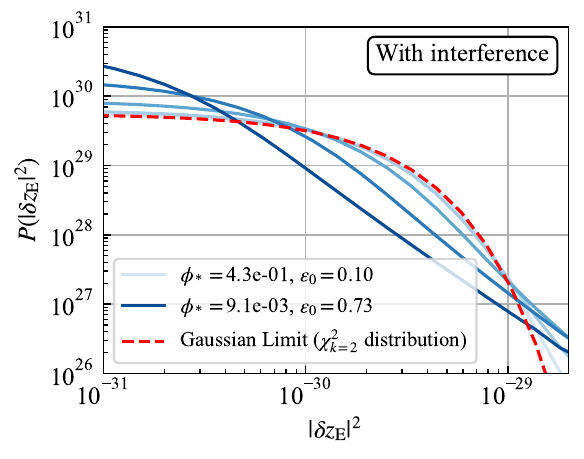}    \includegraphics[width=0.48\linewidth]{ 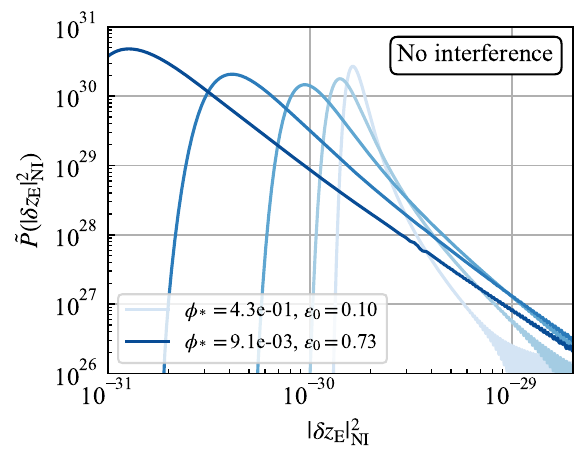}
    \caption{The PDF of $|\delta z_{\rm E}|^2$ and $|\delta z_{\rm E}|^2_{\rm NI}$. We consider $f=10$ nHz in both subplots, with a bandwidth $\Delta f=2$ nHz. Curves of lighter colors correspond to SMBHB populations with more SMBHBs but lighter black hole masses. Note that on the left panel, the ``Gaussian Limit" corresponds to the $\chi^2_k$ distribution with $k=2$, which is also an exponential distribution. }
    \label{fig:PDF1}
\end{figure*}

In comparison, in the Gaussian limit $K_{\delta z}^{\rm Gaus}(t_{\rho}) \propto -t^2_{\rho}/2$, meaning $K^{{\rm Gaus} \prime\prime}_{\delta z}(t_\rho)$ equals a constant. To recover the Gaussian limit, we can artificially rescale $\dd \overline{N}\rightarrow x\,\dd \overline{N}$ and $h^2\rightarrow h^2/x$, and take the limit of $x\rightarrow +\infty$, which corresponds to the limit of an infinite number of arbitrarily weak GW sources. That is to say,
\be \label{eq:var_hc2}
\begin{aligned}
K_{\delta z}^{\rm Gaus \prime\prime}(t_{\rho}) &= K_{\delta z}^{\prime\prime}(0) 
 =-\frac{1}{6} h_c^2 \ \Delta\ln f \ ,
\end{aligned}
\ee 
where we have used Eqs.~(\ref{eq:hc2}),  (\ref{eq:hp22}), and (\ref{eq:P_lambda}).

In the left panel of Fig.~\ref{fig:CGF_Krho}, we show the numerical results of $K^{\prime\prime}(t_\rho)$
for a range of SMBHB populations with different population parameters $(\phi_*, \epsilon_0)$
that are consistent with 15-year NANOGrav data (the black-dashed line in Fig.~\ref{fig:phi0_epsilon0}),
all having the same variance $-K^{\prime\prime}_{\delta z}(0)$ [Eqs.~(\ref{eq:variance}), and (\ref{eq:var_hc2})]. 
As expected, $K^{\prime\prime}(t_\rho)$ is closer to the Gaussian limit for a larger number of GW sources (higher $\phi_*$).
The comparison demonstrates the origin of the non-Gaussian nature of the SGWB from SMBHBs.

The 2D probability distribution, properly normalized, can be expressed in terms of an integral
\begin{equation}
\begin{aligned}
    &P(\delta z_{\rm E,Re},\delta z_{\rm E,Im})\\
    =& \frac{1}{(2\pi)^2}\int_{-\infty}^{\infty} \dd t_{\rm Re}\int_{-\infty}^{\infty} \dd t_{\rm Im} e^{\left[K_{\delta z}(t_{\rho})\right]^*}+it_{\rm Re}\delta z_{\rm E,Re} + it_{\rm Im}\delta z_{\rm E,Im}\\
     =& \frac{1}{2\pi}\int_0^{+\infty} \dd t_{\rho}\,t_{\rho}e^{\left[K_{\delta z}(t_{\rho})\right]^*}J_0\left(t_{\rho}|\delta z_{\rm E}|\right) := f(|\delta z_{\rm E}|)\ .
     \label{eq:PDF_true}
\end{aligned}
\end{equation}

In practice, we define $t_{\rho} = e^{\tau}$ and numerically implement the integral using the $\tau$ variable instead.
It is
important to notice that $\delta z_{\rm E,Re}$ and $\delta z_{\rm E,Im}$ are generally not
statistically independent, i.e. $P(\delta z_{\rm E,Re},\delta z_{\rm E,Im}) \neq P(\delta z_{\rm E,Re}) P(\delta z_{\rm E,Im}) $.  The marginalized distribution of $\delta z_{\rm E,Re}$ is
\begin{equation}
  \begin{aligned}
    P(\delta z_{\rm E,Re})
    =2 \int_0^{+\infty}f( |\delta z_{\rm E}|)\,\dd \sqrt{|\delta z_{\rm E}|^2 -
        \delta z_{\rm E,Re}^2} \ .
    \end{aligned}
  \end{equation}
On the right panel of Fig.~\ref{fig:CGF_Krho}, we show $P(\delta z_{\rm E,Re})$ for several $\phi_*-\epsilon_0$ pairs (same as those in the left panel).
We note that the true distribution 
$P(\delta z_{\rm E,Re})$ is both heavy tailed and strong peaked.

    \begin{figure*}[t]
      \centering
      \includegraphics[width=0.49\linewidth]{ 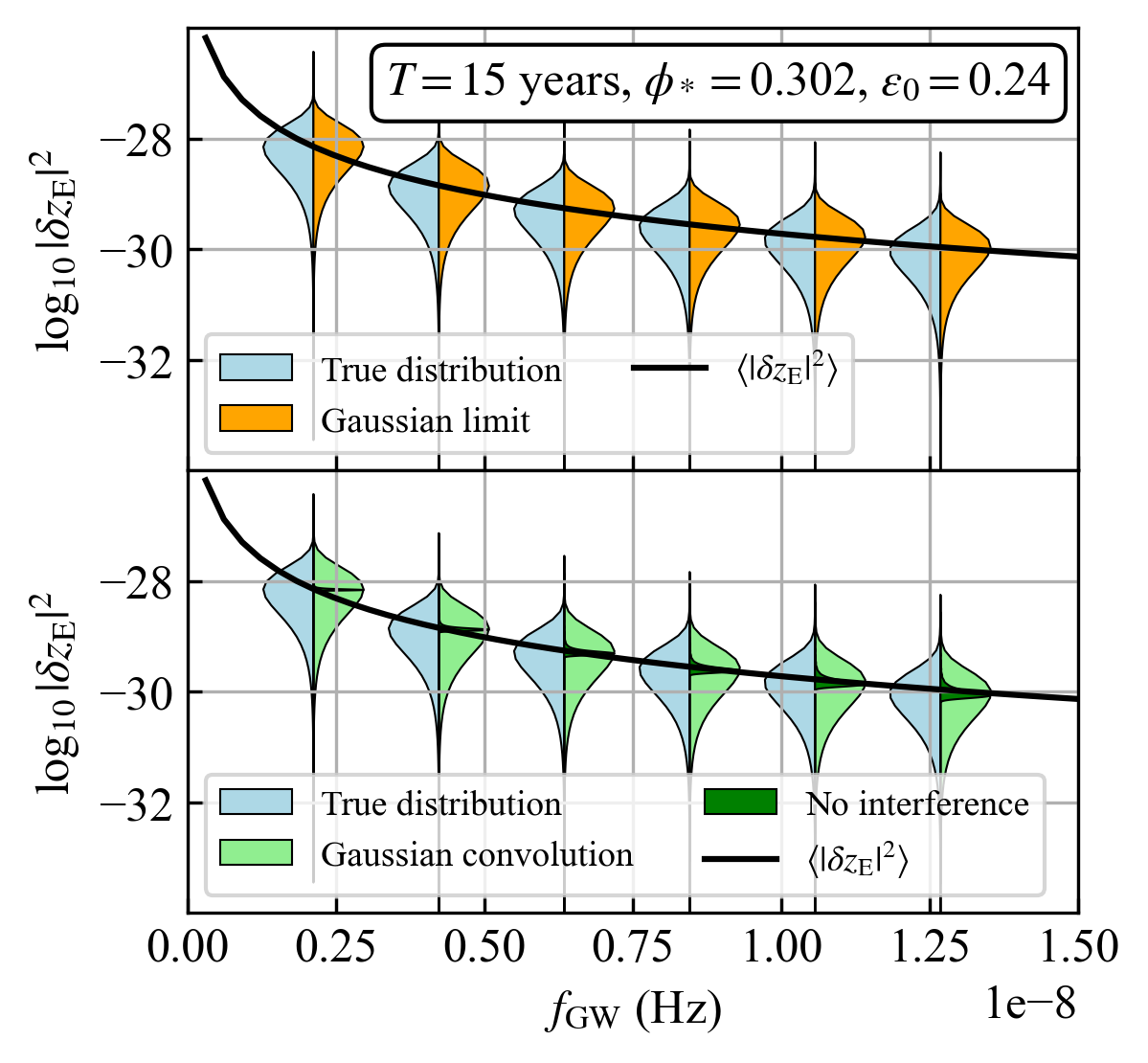}
      \includegraphics[width=0.49\linewidth]{ 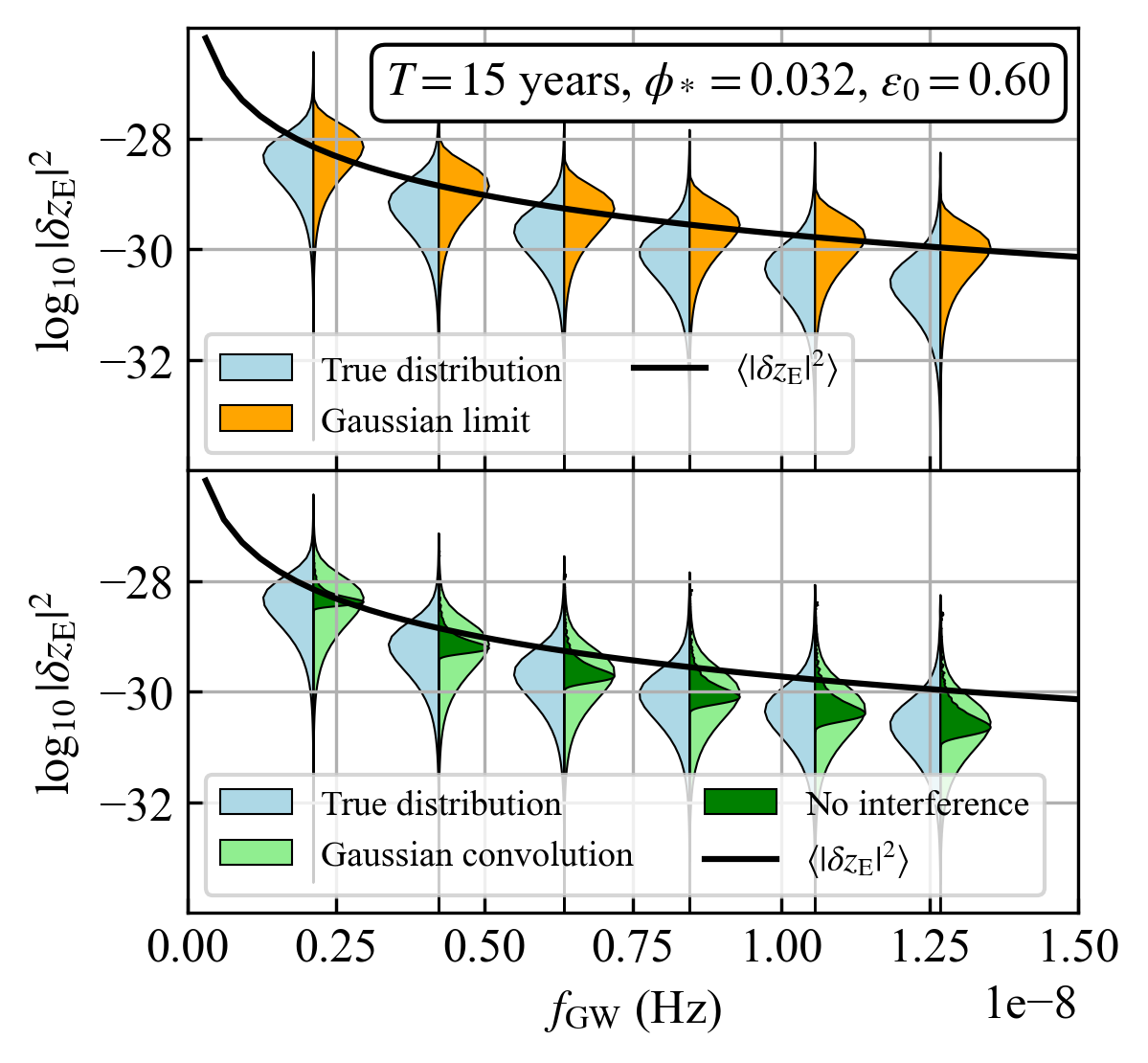}
      \includegraphics[width=0.49\linewidth]{ 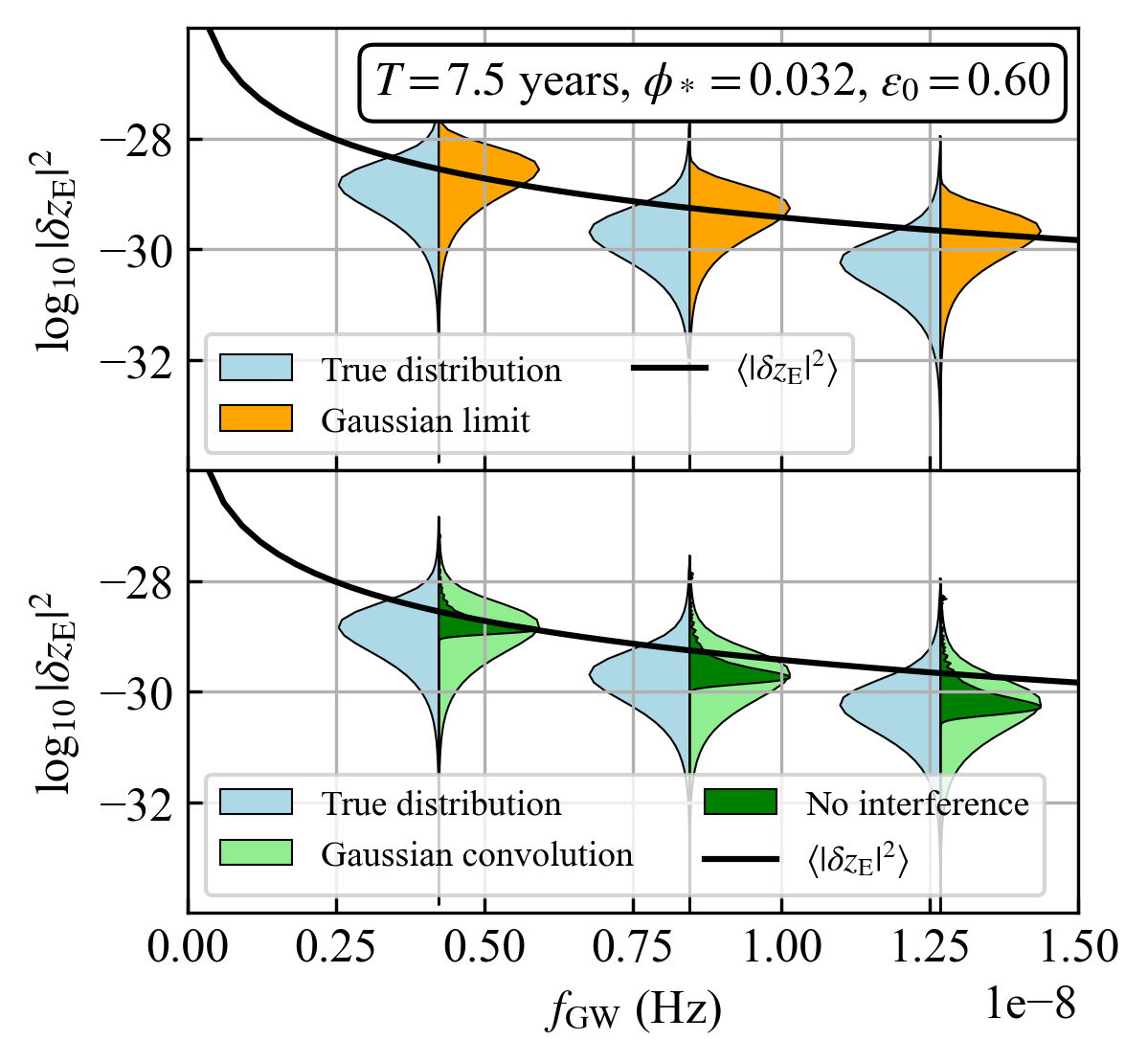}
      \includegraphics[width=0.49\linewidth]{ 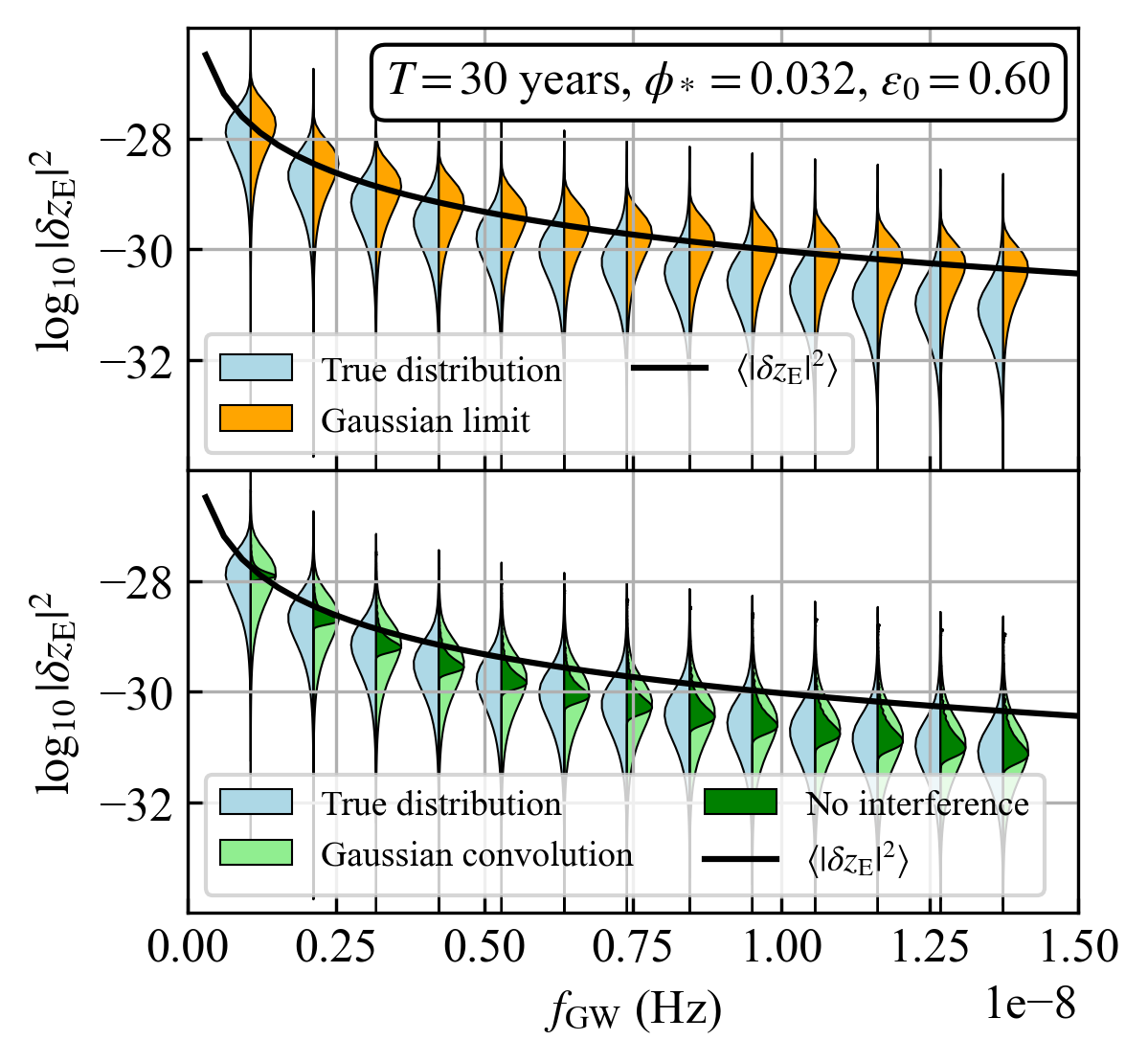}
      \caption{Violin plot for $P(\log_{10}\,|\delta z_{\rm E}|^2)$ and $\tilde{P}(\log_{10}\,|\delta z_{\rm E}|^2_{\rm NI})$ in discrete frequencies binned at a width $\Delta f=1/T$. All violins at the same frequency in each panel share the same expectation value of $\langle|\delta z_{\rm E}|^2\rangle$, which equals to $\langle|\delta z_{\rm E}|^2_{\rm NI}\rangle$, shown by the black curve.
      }
      \label{fig:Violin}
  \end{figure*}

\begin{figure}[t]
    \centering
    \includegraphics[width=0.98\linewidth]{ 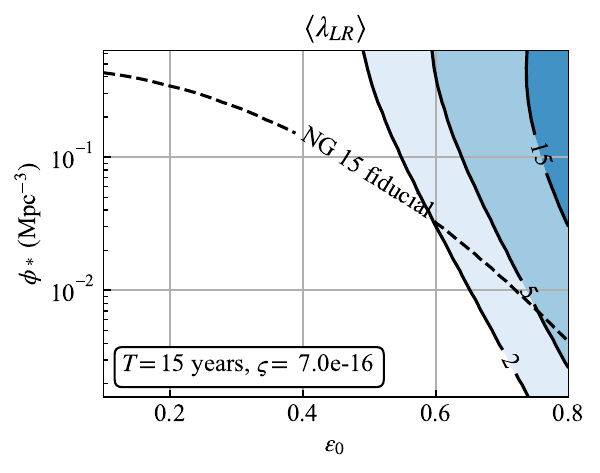}
    \includegraphics[width=0.98\linewidth]{ 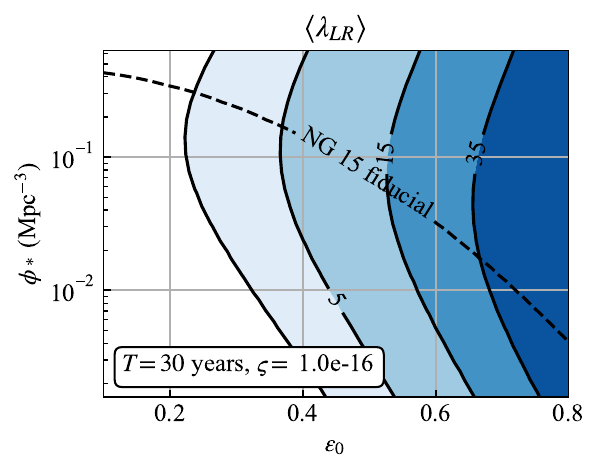}
    \caption{The contours of the expectation value of $\lambda_{\rm LR}$, $\langle\lambda_{\rm LR}\rangle$, on the $\phi_*-\epsilon_0$ plane. The top panel corresponds to the current PTA sensitivity. The statistics $\lambda_{\rm LR}$, defined in Eq.~(\ref{eq:lambda_LR}), characterize the likelihood ratio between a non-Gaussian PDF and the Gaussian PDF, assuming the data is drawn from the non-Gaussian PDF.  }
    \label{fig:lambdaLR}
\end{figure} 
The PDF of $|\delta z_{\rm E}|^2$ is
\begin{equation}
    \begin{aligned}
        P(|\delta z_{\rm E}|^2) = \pi f(|\delta z_{\rm E}|)\ ,
    \end{aligned}
\end{equation}
and is shown in Fig.~\ref{fig:PDF1}.

So far, the calculation in this section is based on Eq.~(\ref{eq:interference}), which includes interference between individual sources. Here, we consider another commonly studied quantity -- the signal power summed incoherently over sources:
\begin{equation}
    \begin{aligned}
        |\delta z_{\rm E}|^2_{\rm NI} \equiv \sum_n\Delta N_n\  \langle|\delta z_{{\rm E}}^{\rm circ}|_n^2\rangle_{\cos\theta,\phi,\cos\iota,\psi,\varphi}\,,
    \end{aligned}
\end{equation}
where the source parameter space is defined only by $\Mc$, $z$ and $f_r$, and $\langle|\delta z_{{\rm E}}^{\rm circ}|^2_n\rangle_{\cos\theta,\phi,\cos\iota,\psi,\varphi} = \frac{2}{15}h_0(f_n,\Mc_n,z_n)^2$, the subscript ``NI" stands for ``No Interference". It is easy to prove that $ \langle |\delta z_{\rm E}|^2_{\rm NI}\rangle = \langle |\delta z_{\rm E}|^2\rangle$. However, comparing to the true $|\delta z_{\rm E}|^2$, cross terms such as $(\delta z_{\rm E}^{\rm circ})_n(\delta z_{\rm E}^{\rm circ})_{n'}$ with $n\neq n'$, which correspond to interference between sources, are absent in the NI counterpart. The corresponding CGF is
  \begin{equation}
      \begin{aligned}
          \tilde{K}_{|\delta z|^2}(t) \equiv &\Delta\ln f\iint \dd \log_{10}\Mc\dd z 
   \\
   &\left[\exp\left(\frac{2}{15}\,i\,t\,h_0^2 \right) -1 \right]\frac{\dd^3 \overline{N}}{\dd \log_{10}\Mc\,\dd z\, \dd \ln f_r} \ ,\label{eq:K_woIF}
      \end{aligned}
  \end{equation}
and the PDF is
\begin{equation}
    \begin{aligned}
        \tilde{P}(|\delta z_{\rm E}|^2_{\rm NI}) =\frac{1}{2\pi}\int_{-\infty}^{+\infty}  \dd t\, e^{\left[\tilde{K}_{|\delta z|^2}(t)\right]^*}+i \,t\, |\delta z_{\rm E}|^2_{\rm NI} \ .\label{eq:PDF_nointerference}
    \end{aligned}
\end{equation}

In Fig.~\ref{fig:PDF1}, we find that the tails of $P(|\delta z_{\rm E}|^2) $ and $\tilde{P}(|\delta z_{\rm E}|^2_{\rm NI}) $ can be very well approximated by power law distributions. This result is in agreement with the findings of Refs.~\cite{Ellis:2023owy,Ellis:2023dgf,Raidal:2024odr}. The major difference is that in the no interference case, $\tilde{P}(|\delta z_{\rm E}|^2_{\rm NI})$ is strong-peaked at some positive value 
while the peak is smeared and always located at zero in the presence of interference. 

In Fig.~\ref{fig:Violin}, we show the PDF of $\log_{10} |\delta z_{\rm E}|^2$ 
\begin{equation}
    \begin{aligned}
        P(\log_{10} |\delta z_{\rm E}|^2) = \ln 10\,|\delta z_{\rm E}|^2\,P( |\delta z_{\rm E}|^2)\ ,
    \end{aligned}
\end{equation}
and the no interference counterpart $\tilde{P}(\log_{10} |\delta z_{\rm E}|^2_{\rm NI})$
at different frequency bins for two SMBHB populations and three total observation times.
The true PDF is much more diffused than the no interference case.
The comparison shows that interference dominates the variance in $|\delta z_{\rm E}|^2$, while Poisson fluctuations play a minor role.
In the no interference case, the expectation value of $|\delta z_{\rm E}|^2_{\rm NI}$ is consistently higher than the median of $|\delta z_{\rm E}|^2_{\rm NI}$, especially at higher frequencies. 
This is because $\tilde{P}(\log_{10}|\delta z_{\rm E}|^2_{\rm NI})$ is a skewed distribution that decreases to zero rather slowly at large $|\delta z_{\rm E}|^2_{\rm NI}$, contrary to rapid fall-off of the low $|\delta z_{\rm E}|^2_{\rm NI}$ end (Fig.~\ref{fig:PDF1}). This observation is consistent with the Monte-Carlo results by Refs.~\cite{sesana2008stochastic, kelley2017massive}. The difference between the median and mean values is more prominent for combinations between smaller $\phi_*$ with larger $\epsilon_0$, which correspond to stronger non-Gaussianity. 
We also find that the true PDF with the interference effect is well approximated by a Gaussian convolution of the no interference PDF
\begin{equation}
    \begin{aligned}
        \tilde{P}(\delta z_{\rm E,Re},\delta z_{\rm E,Im}) &= \int_{0}^{+\infty}\dd |\delta z_{\rm E}|^2_{\rm NI} \,\tilde{P}(|\delta z_{\rm E}|^2_{\rm NI})\,\frac{ e^{-(\delta z_{\rm E,Re}^2 + \delta z_{\rm E,Im}^2)/|\delta z_{\rm E}|^2_{\rm NI}}}{\pi\,|\delta z_{\rm E}|^2_{\rm NI} }\ ,\label{eq:PDF_convolution}
    \end{aligned}
\end{equation}
where the numerical comparison between $P(\delta z_{\rm E,Re},\delta z_{\rm E,Im})$ and $\tilde{P}(\delta z_{\rm E,Re},\delta z_{\rm E,Im})$ is shown in Fig.~\ref{fig:Violin}. Numerically, the two distributions are nearly indistinguishable from the realistic SMBHB population models we consider. However, in Appendix~\ref{appendix}, we demonstrate that they are not mathematically identical. 

Eq.~(\ref{eq:PDF_convolution}) as an excellent approximate PDF implies that non-Gaussianity originates from the Poisson fluctuation of the source numbers. 
The interference effect is well approximated by randomly drawing the real and imaginary parts of $\delta z_{\rm E}$ from a 2D Gaussian distribution. It justifies the population inference done in \cite{Sato-Polito:2024lew} at the practical level in the following sense:
NANOGrav had first derived the likelihoods for the frequency-binned power $|\delta z_{\rm E}|^2_{\rm NI}$ from the redshift residuals of individual pulsars under the assumption that the SGWB is a Gaussian random field~\cite{NANOGrav:2023hfp}. \cite{Sato-Polito:2024lew} then used these likelihoods as summary data and performed Bayesian population inference using the correct non-Gaussian PDF $\tilde P(|\delta z_{\rm E}|^2_{\rm NI})$ for the total power (hence without inference). These two analyses combined are equivalent to using Eq.~(\ref{eq:PDF_convolution}) to perform Bayesian inference at the redshift observable level.

In principle, one would like to perform Bayesian inference for SMBHB population models by applying a multi-variate non-Gaussian PDF $P(\{\delta z^{(I)}_{\rm E, Re},\delta z^{(I)}_{\rm E, Im} \})$ for the redshifts of all $N_p$ pulsars $\delta z^{(I)}_{\rm E},\,I=1,2,\cdots,N_p$ to direct PTA observables. Unfortunately, writing down an exact expression for this PDF seems prohibitive\footnote{Even though a straightforward generalization of the formalism we present in this work will allow the calculation for the corresponding multi-variate characteristic function, inverse Fourier transform in high dimensions would be impractical.}, as redshift signals in different pulsars are correlated in the Earth term. 

However, the numerical success of the approximate PDF in Eq.~(\ref{eq:PDF_convolution}) guides us to conjecture that if one convolves $\tilde P(|\delta z_{\rm E}|^2_{\rm NI})$ with a multi-variate Gaussian distribution for the redshifts of all pulsars, with the correct covariance set up between pulsars:
\begin{equation}
    \begin{aligned}
        \tilde{P}(\{\delta z^{(I)}_{\rm E, Re},\delta z^{(I)}_{\rm E, Im} \}) &= \int_{0}^{+\infty}\dd |\delta z_{\rm E}|^2_{\rm NI} \,\tilde{P}(|\delta z_{\rm E}|^2_{\rm NI})\,\frac{ e^{-\boldsymbol{\delta z}_{\rm E}^\dagger\,\boldsymbol{C}^{-1}(|\delta z_{\rm E}|^2_{\rm NI})\,\boldsymbol{\delta z}_{\rm E}}}{\pi^{N_p}\,{\rm det}\,\boldsymbol{C}(|\delta z_{\rm E}|^2_{\rm NI})}\ ,
        \label{eq:PDF_convolution_multipulsar}
    \end{aligned}
\end{equation}
the result may very well approximate the exact joint PDF. In Eq.~(\ref{eq:PDF_convolution_multipulsar}), the vector $\boldsymbol{\delta z}_{\rm E}$ collectively represents redshifts of all pulsars, and $\boldsymbol{C}(|\delta z_{\rm E}|^2_{\rm NI})$ stands for the multi-pulsar covariance matrix that would result from an isotropic Gaussian SGWB with a given full-sky average power $|\delta z_{\rm E}|^2_{\rm NI}$. We note that applying Eq.~(\ref{eq:PDF_convolution_multipulsar}) to Bayesian inference at the level of individual pulsar redshifts will be mathematically equivalent to the combination of two steps: first deriving the likelihoods for the mean frequency-binned powers assuming a Gaussian random SGWB, and then inferring the SMBHB population by applying $\tilde{P}(|\delta z_{\rm E}|^2_{\rm NI})$ to the derived likelihoods as summary data. 

If true, such an approach will be a practical path toward full Bayesian treatment, including non-Gaussianity and at the level of pulsar redshifts. However, how this idea may be quantitatively validated is an open question.

\subsection{Sensitivity estimation}
\label{subsec:test}

With the non-Gaussian statistics developed above, we are in a position to estimate the PTA sensitivities in measuring the non-Gaussianity in the nHz SGWB and constraining the SMBHB population parameters and to test possible bias in inferring the population properties if the Gaussian statistics are applied instead.

\subsubsection{Likelihood ratio between non-Gaussian and Gaussian model}
We first implement a simple likelihood ratio test by applying both the non-Gaussian and the Gaussian statistics to mock data generated from a SMBHB population model. 
We assume that measurement noises for $\delta z_{\rm Re}$ and $\delta z_{\rm Im}$ follow a Gaussian distribution with the variance of $\varsigma$. In the presence of noise, the new PDF
$P(\delta z_{\rm E,Re} , \delta z_{\rm E,Im}|\varsigma)$ is calculated in the similar way, see Eq.~(\ref{eq:PDF_true}), with
the CGF modified as 
\begin{equation}
    \begin{aligned}
         K_{\delta z}(t_{\rho}|\varsigma) = K_{\delta z}(t_{\rho})  -\frac{\varsigma^2 t_{\rho}^2}{2}\ .
    \end{aligned}
\end{equation}
For comparison, 
the CGF for a Gaussian PDF $P_{\rm Gaus}(\delta z_{\rm E,Re} , \delta z_{\rm E,Im}|\varsigma)$ with the same variance is
\begin{equation}
    \begin{aligned}
        K_{\delta z}^{\rm Gaus}(t_{\rho}|\varsigma) = \frac{t_{\rho}^2 K_{\delta z}^{\prime\prime}(0|\varsigma)}{2}\ .
    \end{aligned}
\end{equation}

We benchmark two observational noise levels: one with a noise level $\varsigma^k = 7 \times 10^{-15}$ over $T = 15$ years, and another with a noise level $\varsigma^k = 1 \times 10^{-15}$ over $T = 30$ years, where $k$ labels the $k$-th frequency bin. The first benchmark case is consistent with the current PTA sensitivity, where the SGWB signal stands out of noise in five frequency bins.

We define the likelihood ratio test statistic as follows:
\begin{equation}
    \begin{aligned}
        \lambda_{\rm LR} =  2\ln \frac{P(\boldsymbol{\delta z}_{\rm E,Re},\boldsymbol{\delta z}_{\rm E,Im}|\boldsymbol{\varsigma})}{ P_{\rm Gaus}(\boldsymbol{\delta z}_{\rm E,Re},\boldsymbol{\delta z}_{\rm E,Im}|\boldsymbol{\varsigma})}\ ,\label{eq:lambda_LR}
    \end{aligned}
\end{equation}
 where $P(\boldsymbol{\delta z}_{\rm E,Re},\boldsymbol{\delta z}_{\rm E,Im}|\boldsymbol{\varsigma})=\prod_{k=1}^{k_{\rm max}} P(\delta z_{\rm E,Re}^{k},\delta z_{\rm E,Im}^{k}|\varsigma^k)$ and $\delta z_{\rm E,Re}^k$ and $\delta z_{\rm E,Im}^k$ are the real and imaginary components of the redshift data in the $k$-th frequency bin. 
We ensure that the Gaussian distribution, $P_{\rm Gaus}(\delta z_{\rm E,Re}^k, \delta z_{\rm E,Im}^k | \varsigma^k)$, has the same variance as the non-Gaussian distribution $P(\delta z_{\rm E,Re}^k, \delta z_{\rm E,Im}^k | \varsigma^k)$. We then evaluate the expectation value $\langle \lambda_{\rm LR} \rangle$ using the following formula
\begin{equation}
    \begin{aligned}
        \langle\lambda_{\rm LR} \rangle= \int \dd \boldsymbol{\delta z}_{\rm E,Re} \dd \boldsymbol{\delta z}_{\rm E,Im}\,P(\boldsymbol{\delta z}_{\rm E,Re},\boldsymbol{\delta z}_{\rm E,Im}|\boldsymbol{\varsigma})\lambda_{\rm LR}\ .
    \end{aligned}
\end{equation}
We choose $k_{\rm max} = 14$ and $k_{\rm max} = 28$ for the two benchmark cases described above. The integral can be simplified using the fact that all frequency bins are independent of each other, and that at each bin the PDFs only have dependence on the modulus of $\delta z_{\rm E}^k$.

The results are shown in Fig.~\ref{fig:lambdaLR}. With current PTA sensitivity (which has five frequency bins above the noise floor), we expect to find $\langle \lambda_{\rm LR} \rangle > 2$ for $\epsilon_0 > 0.6$. This may increase to $\langle \lambda_{\rm LR} \rangle > 15$ for the same parameter range in the more optimistic situation.

\begin{figure}[t]
    \centering
\includegraphics[width=0.98\linewidth]{ 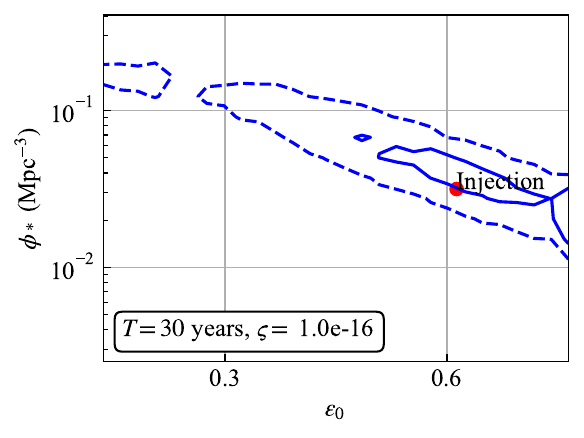}
    
    \caption{
    The contour of the best-fit parameters using the non-Gaussian PDF we derived in this work. We choose one pair of the population model parameters $(\phi_*,\epsilon_0)$ and simulate 10,000 realizations of mock data. The true parameters are marked as the red point. We then calculate the likelihood values for each realization using the non-Gaussian PDF and investigate the distributions of the best-fit points. The solid and dashed contours indicate $1\sigma$ and $2\sigma$ confidence regions, respectively. 
    }
    \label{fig:parameter_estimation}
\end{figure}

\begin{figure}[t]
    \centering
    \includegraphics[width=0.98\linewidth]{ 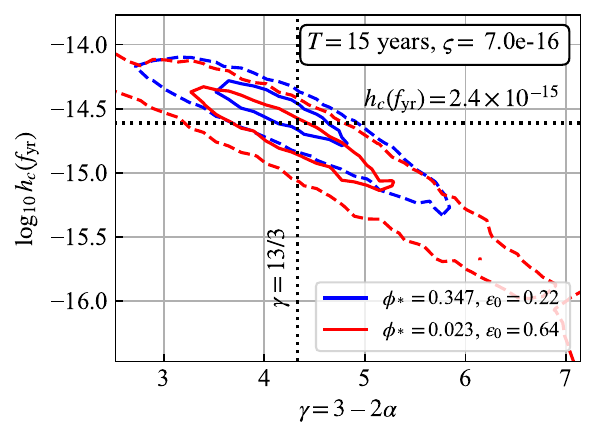}
    \includegraphics[width=0.98\linewidth]{ 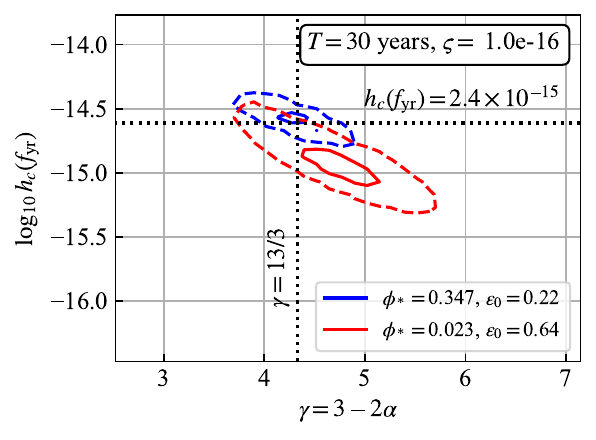}
    \caption{The contours of the best-fit parameters using the classical power-law model. We assume two different population model parameters and simulate 200,000 realizations of mock data for each. We then calculate the likelihood values for each realization using the power-law model and investigate the distributions of the best-fit points. The solid and dashed contours indicate $1\sigma$ and $2\sigma$ confidence regions, respectively. }
    \label{fig:PLfit}
\end{figure}

\subsubsection{Parameter estimation of the SMBHB population model}
We also perform a Monte Carlo simulation, in which we generate random data points drawing from the non-Gaussian PDF for a given combination of $\phi_*$ and $\epsilon_0$.

We then define the likelihood function as follows
\begin{equation}
    \mathcal{L}(\log_{10}\phi_*,\epsilon_0|\boldsymbol{\delta z}_{\rm E, Re}^{\rm mock},\boldsymbol{\delta z}_{\rm E, Im}^{\rm mock}) = P(\boldsymbol{\delta z}_{\rm E,Re}^{\rm mock},\boldsymbol{\delta z}_{\rm E,Im}^{\rm mock}|\boldsymbol{\varsigma},\phi_*,\epsilon_0)\ ,
\end{equation}
where $P(\boldsymbol{\delta z}_{\rm E,Re}^{\rm mock},\boldsymbol{\delta z}_{\rm E,Im}^{\rm mock}|\boldsymbol{\varsigma},\phi_*,\epsilon_0)$ is the calculated with the population model parameters $\phi_*$, $\epsilon_0$, while other parameters are fixed at their fiducial values.

Next, we choose the second benchmark and use the population model parameters $(\phi_*,\epsilon_0)=(0.032, 0.61)$ to generate mock data ($\boldsymbol{\delta z}_{\rm E,Re}^{\rm mock},\boldsymbol{\delta z}_{\rm E,Im}^{\rm mock}$) 10,000 times. Then for each realization, we calculate the likelihood values on the $\phi_*$--$\epsilon_0$ plane and find the maximal-likelihood value points. We then collect the best-fit points and investigate their distribution. The results are shown in Fig.~\ref{fig:parameter_estimation}.

\subsubsection{Strain power spectrum inference}
PTA data have been used in inferring the strain power spectrum of the nHz SGWB, assuming the strain is a Gaussian random field with a power-law frequency dependence 
\be 
h_c^2(f) = h_c^2(f_{\rm yr})(f/f_{\rm yr})^{2\alpha}\ .
\ee 
In this subsection, we apply the Gaussian statistics to mock data generated from a SMBHB population model and quantify the resulted bias.
We construct the log-likelihood function as follows
\begin{equation}
    \begin{aligned}
        &\ln\mathcal{L}(\boldsymbol{\delta z}_{\rm E, Re}^{\rm mock},\boldsymbol{\delta z}_{\rm E,Im}^{\rm mock}|\log_{10}h_c(f_{\rm yr}),\gamma)\\
        =& \sum_k\left[- |\delta z_{\rm E}^k|^2 / \langle |\delta z_{\rm E}^k|^2\rangle -\ln \pi\ \langle |\delta z_{\rm E}^k|^2\rangle  \right]\ ,
    \end{aligned}
\end{equation}
where 
\begin{equation}
    \begin{aligned}
\langle|\delta z_{\rm E}^k|^2\rangle = \frac{1}{3}h_c^2(f_k)\Delta \ln f_k + 2(\varsigma^k)^2.
    \end{aligned}
\end{equation}
In accordance with the convention of PTA community,  we have introduced the notation $\gamma$,
which is the power index of the time delay power spectrum and is related to $\alpha$ by $\alpha = (3-\gamma)/2$. Similarly, we choose the earlier defined two benchmarks and simulate 200,000 random realizations for given population model parameters, then collect the best-fit $(\log_{10}h_c(f_{\rm yr}),\,\gamma)$ points, and plot the distribution. We show the results in Fig.~\ref{fig:PLfit}. 
Comparing the blue and red contours, we find that as the non-Gaussianity increases, the characteristic strain $h_c(f_{\rm yr})$ tends to be underestimated. At the same time, the power-law index $\gamma$ is overestimated, consistent with what is implied in Fig.~\ref{fig:Violin}. This observation is in agreement with findings in \cite{Becsy:2023qul}. The Gaussian statistics is still a good approximation in inferring the strain power spectrum with current PTA sensitivity and will bias the inference in the foreseeable future.

\bigskip

In conclusion, with the current sensitivity, PTAs cannot determine SMBHB population model parameters with a decent precision.
This is fundamentally limited by the large variance arising from interference and the limited number of frequency bins above the noise level. However, as observation period increases and pulsar timing quality improves, it will become possible to constrain these parameters more precisely in the foreseeable future.

\section{Conclusions}\label{sec:conclusions}

Since the PTA collaborations dropped the first evidence for the nanohertz SGWB, there has been an ongoing debate over whether it originates from astrophysical sources, such as SMBHBs, or has a primordial origin.
These two broad hypotheses suggest different models for constructing the SGWB observed today: fewer sources with stronger individual signals or more sources with weaker individual signals. In this work, we have developed a semianalytic mathematical framework for computing the non-Gaussian PDF of the redshift $P(\delta z)$ for a SMBHB population model,
where we have accounted for both Poissonian fluctuations in the number
of SMBHBs and GW interference.

To quantify the significance of this distinction, we have numerically calculated the exact PDF of the GW strain power in the frequency domain as a function of population model parameters. With current PTA sensitivities, evidence of non-Gaussianity may be detected in some areas of the population parameter space with fewer and individually louder SMBHBs. As PTA sensitivity improves over time, we expect more robust evidence for non-Gaussianity. We also find the Gaussian statistics are still a good approximation in
inferring the strain power spectrum $h_c^2(f)$ with current PTA sensitivity, though it will bias the inference as PTA data of lower noise 
accumulates in the foreseeable future.

We have proposed an approximated formula to calculate the PDF incorporating data from many pulsars across the sky. We have shed light on why the approximation is numerically very close to the exact answer while mathematically not the same. The correction of the approximated PDF will naturally lead to modifications in two-point correlation functions \cite{Allen:2024rqk,Allen:2024mtn,Bernardo:2024uiq,Bernardo:2024bdc,Konstandin:2024fyo,Wu:2024xkp}. Additionally, developing a numerically efficient method to calculate the analytical PDFs for data analysis will be beneficial. These topics will be addressed in future work.

\acknowledgements
We thank Reginald Christian Bernardo, Luke Zoltan Kelley, Thomas Konstandin, Enrico Perboni, Gabriela Sato-Polito, Ye-Fei Yuan, and Matias Zaldarriaga for useful discussions.

IFAE is partially funded by the CERCA program of the Generalitat de Catalunya. X.X. is partly funded by the Grant No. CNS2023-143767 and Grant No. CNS2023-143767 funded by MICIU/AEI/10.13039/501100011033 and by European Union NextGenerationEU/PRTR.
X.X. is supported by Deutsche
Forschungsgemeinschaft under Germany’s Excellence Strategy EXC2121 “Quantum Universe” — 390833306.  L.D. acknowledges research grant support from the Alfred P. Sloan Foundation (Award No. FG-2021-16495), and support of Frank and Karen Dabby STEM Fund in the Society of Hellman Fellows.

\appendix
\section{Toy model}\label{appendix}
To see if the approximation in Eq.~(\ref{eq:PDF_convolution}) works, we perform Fourier transform of $\tilde{P}(\delta z_{\rm E,Re},\delta z_{\rm E,Im})$ defined in Eq.~(\ref{eq:PDF_convolution}) to obtain the conjugate of the characteristic function (CF)
\begin{equation}
    \begin{aligned}
    \tilde{\Phi}^*(t_{\rho}) = & \int_{-\infty}^{+\infty} \dd\delta z_{\rm Re}\int_{-\infty}^{+\infty} \dd \delta z_{\rm Im}\, \\
    &\qquad\qquad\tilde{P}(\delta z_{\rm E,Re},\delta z_{\rm E,Im})\, e^{-i \,t_{\rm Re}\,\delta z_{\rm Re}  -  i\,t_{\rm Im}\,\delta z_{\rm Im}}
    \\
         = &\int_{0}^{+\infty} \dd |\delta z_{\rm E}|_{\rm NI}^2 \,\tilde{P}(|\delta z_{\rm E}|^2_{\rm NI})\,\exp\left(-
         t_{\rho}^2\,|\delta z_{\rm E}|^2_{\rm NI}/4
         \right)\ ,\label{appeq:convolution}
    \end{aligned}
\end{equation}
where CGF is the logarithm of the CF
\begin{equation}
    \begin{aligned}
        \tilde{\Phi}(t_\rho) \equiv \exp \tilde{K}_{\delta z}(t_{\rho}),\qquad \tilde{K}_{\delta z}(t_{\rho}) \equiv \ln\tilde{\Phi}(t_\rho)\,.
    \end{aligned}
\end{equation}
As a simple example, we consider a binary population where the binary number density is a delta function of  $\log_{10}\Mc$, $z$ and $\ln f_r$
\begin{equation}
    \begin{aligned}
        &\frac{\dd^3 \overline{N}}{\dd \log_{10}\Mc \,\dd z\,\dd\ln f_r} \\
        =&\overline{N} \,\delta(\log_{10}\Mc - \log_{10}\Mc_0)\,\delta(z - z_0)\delta(\ln f_r-\ln f_{r,0}).
    \end{aligned}
\end{equation}
Inserting the above population model into Eq.~(\ref{eq:K_woIF}) and (\ref{eq:PDF_nointerference}) we find the PDF as follows
\begin{equation}
    \begin{aligned}
        \tilde{P}(|\delta z_{\rm E}|^2_{\rm NI}) = e^{-\overline{N}}\sum_{k=0}^{+\infty}\frac{\overline{N}^k}{k!}\delta\left(|\delta z_{\rm E}|^2_{\rm NI} - \frac{2}{15}k\,h_0^2\right)\ ,
    \end{aligned}
\end{equation}
here $k$ are integers, $h_0 = h_0(f_0,\Mc_0,z_0)$, and $f_0=f_{r,0}/(1+z_0)$. Inserting the above expression in Eq.~(\ref{appeq:convolution}), we easily find the result for the CF 
\begin{equation}
    \begin{aligned}
        \tilde{\Phi}^*(t_\rho) &= e^{-\overline{N}}\sum_{k=0}^{+\infty}\frac{\overline{N}^k}{k!}\exp\left(-
         \frac{1}{30}k\,t_{\rho}^2\,h_0^2
         \right)\\
         & = \exp\left\{
         \left[\exp\left(-\frac{1}{30}t_\rho^2\,h_0^2 \right) -1 \right]\overline{N}
         \right\}\ .
    \end{aligned}
\end{equation}
The CGF of the approximated PDF is
\begin{equation}
    \begin{aligned}
        \tilde{K}_{\delta z}(t_{\rho}) = \ln\tilde{\Phi}(t_\rho)= \left[\exp\left(-\frac{1}{30}t_\rho^2\,h_0^2 \right) -1 \right]\overline{N}\ .\label{appeq:K_approx}
    \end{aligned}
\end{equation}

\begin{figure}[h]
    \centering
    \includegraphics[width=0.9\linewidth]{ 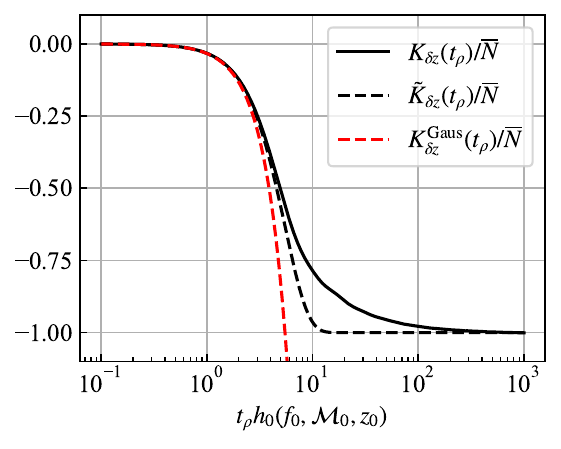}
    \caption{We compare $K_{\delta z}(t_\rho)/\overline{N}$, $\tilde{K}_{\delta z}(t_\rho)/\overline{N}$ and $K^{\rm Gaus}_{\delta z}(t_{\rho})/\overline{N}$. We calculate the CGFs by assuming all sources have the same chirp mass $\Mc_0$ and located at the same redshift $z_0$. }
    \label{appfig:CGF_comparison}
\end{figure}

In comparison, the CGF of the exact PDF is
\begin{equation}
    \begin{aligned}
        K_{\delta z}(t_{\rho}) =\int \dd\lambda_zP(\lambda_z)\left[\cos\left( t_{\rho}h_0\lambda_z \right) -1 \right]\overline{N}\ .\label{appeq:K_exact}
    \end{aligned}
\end{equation}
Using Eq.~(\ref{eq:P_lambda}), we easily find
\begin{equation}
    \begin{aligned}
        K^{\prime\prime}_{\delta z}(0) = \tilde{K}^{\prime\prime}_{\delta z}(0) = -\frac{h_0^2\,\overline{N}}{15}\ .
    \end{aligned}
\end{equation}
The difference emerges at the fourth-order cumulant (Kurtosis)
\begin{equation}
    \begin{aligned}
        K^{(4)}_{\delta z}(0) = \frac{3\,h_0^4\,\overline{N}}{175}\ ,\qquad 
        \tilde{K}^{(4)}_{\delta z}(0) = \frac{h_0^4\,\overline{N}}{75}\ .
    \end{aligned}
\end{equation}
In Fig.~\ref{appfig:CGF_comparison}, we compare $\tilde{K}_{\delta z}(t_\rho)$ and $K_{\delta z}(t_\rho)$ in Eq.~(\ref{appeq:K_approx}) and Eq.~(\ref{appeq:K_exact}) with the CDF of the Gaussian distribution, $K^{\rm Gaus}_{\delta z}(t_{\rho}) = -t_{\rho}^2\,h_0^2\,\overline{N}/30$. The ``No interference + Gaussian convolution" approximation correctly captures the asymptotic behavior of the true CGF at $t_{\rho}\rightarrow 0$ and $t_{\rho}\rightarrow +\infty$. However, in the intermediate region, the approximation deviates from the true PDF and underestimates the non-Gaussianity since it is closer to the Gaussian limit.

\bibliography{ref}

\end{document}